\documentclass[12pt, draftcls, onecolumn]{IEEETran}
\renewcommand{\baselinestretch}{2}

\usepackage{amsmath}
\usepackage{amsfonts}
\usepackage{epsfig}
\usepackage{amssymb}

\begin{document}

\title{\huge{Performance Analysis of Sequential Method for Handover in Cognitive Radio Networks}}

\author{\authorblockN{Hossein Shokri-Ghadikolaei, Mohammad Mozaffari,
Masoumeh~Nasiri-Kenari \\{\small Wireless Research Lab., Elec. Eng. Dept., Sharif University of Technology}}}

\maketitle
\begin{abstract}
Powerful spectrum handover schemes enable cognitive radios (CRs)
to use transmission opportunities in primary users' channels
appropriately. In this paper, we consider the cognitive access of
primary channels by a secondary user (SU). We evaluate the average
detection time and the maximum achievable average throughput of
the SU when the sequential method for hand-over (SMHO) is used. We
assume that a prior knowledge of the primary users' presence and
absence probabilities are available. In order to investigate the
maximum achievable throughput of the SU, we end into an
optimization problem, in which the optimum value of sensing time
must be selected. In our optimization problem, we take into
account the spectrum hand over due to false detection of the
primary user. We also propose a weighted based hand-over (WBHO)
scheme in which the impacts of channels conditions and primary
users' presence probability are considered. This spectrum handover
scheme provides higher average throughput for the SU compared to the SMHO
method. The tradeoff between the maximum achievable throughput and
consumed energy is discussed, and finally an energy efficient
optimization formulation for finding a proper sensing time is
provided.
\end{abstract}


\section{Introduction}
\PARstart{E}{merging} new wireless applications and ever-growing
need to have a higher data rate have increased the demand for
accessing to the spectrum in the past ten years, incredibly.
Though the available spectrum resources seem not to meet the
ever-growing demand, many investigations reveal that the spectrum
is inefficiently used \cite{akylidiz}. Cognitive radio (CR)
concept has been introduced to improve spectrum efficiency by
allowing the low-priority secondary users (SUs) to
opportunistically exploit the unused licensed spectrum of the
high-priority primary users (PUs) \cite{akylidiz}, \cite{haykin}.
To this end, first the spectrum holes must be found through
appropriate and reliable spectrum sensing techniques. However,
there are two challenges associated with spectrum sensing: (1)
Limited Observations, and, (2) Time Variation. Since the numbers
of samples used for sensing are limited, the idle spectrum cannot
be detected perfectly. Moreover, due to stochastic nature of PUs
activities, accessibility of the SUs to the spectrum is time
variant. However, the SU enforce to stop its transmission and
vacate the occupied channel when the PU has data to transmit on
this channel. Within the transmission period of a secondary
connection, it is likely to have multiple spectrum handoffs due to
interruptions from the PUs. In order to provide reliable
transmission for the SUs and guarantee some level of quality of
service (QoS), a set of procedures called spectrum handover (HO)
is initiated to help the SU to find a new transmission opportunity
and resume its unfinished transmission \cite{akylidiz},
\cite{lcwang2008}.

Generally, there exists more than one channel to be sensed by a
CR. To deal with this fact, sensing schemes are commonly divided
into two categories, i.e., wideband sensing and narrowband
sensing. Sensing is wideband when multiple channels are sensed
simultaneously. These multiple sensed channels can cover either
the whole or a portion of the primary channels \cite{pedram2011}.
On the other hand, when only one channel is sensed at a time, the
sensing process is narrowband. Ease of implementation, lower power
consumption, and less computational complexity leads to great
interest in narrowband sensing. When the narrowband sensing is
used, the channels have to be sensed in a proper order called
sensing sequence. Incorporating powerful spectrum sensing schemes
enable SUs to transmit and/or receive data while no channels are
dedicated to them. Average throughput of the SU, average sensing
time, and consumed energy are some common metrics considered in
designing appropriate sensing schemes.

Throughput maximization of the SUs has been widely investigated in
the literature. Specifically, in \cite{win2008c} and
\cite{Hamdaoui09} a set of procedures is proposed to determine the
optimal set of candidate channels, and maximizing the spectrum
accessibility through the optimal number of candidate channels is
investigated. Minimizing the overall system time of a SU through
load balancing in probability-based and sensing-based spectrum
decision schemes is investigated in \cite{wangjsac2011}. The joint
design of sensing-channel selection and power control scheme is
investigated in \cite{chung2009}. Assuming a perfect sensing
scheme and static wireless channels, this joint optimization is
formulated, and suboptimal algorithms with tolerable computational
complexity are developed to approximately solve the derived
optimization problem. The same problem is formulated in
\cite{Wang2011} considering the impact of time-varying fading
channels as well as the sensing errors. The author derives a
closed-form relation under the constraint of average available
power and the level of collision with the PUs and develops a
stochastic optimization approach. Throughput maximization through
optimizing spectrum sensing time has gained a lot of interest.
Spectrum sensing time is one of the most effective factors which
must be determined carefully to obtain a powerful sensing scheme.
In \cite{Kim08}, \cite{liang}, \cite{yuan2010}, and \cite{lee} the
impact of spectrum sensing time on the overall throughput of the
SUs is investigated. It is shown in \cite{liang} that as the
sensing time increases, the sensing accuracy increases as well,
but the throughput decreases; thus there is an interesting
tradeoff. In \cite{liang}, while the spectrum HOs effect has not
been taken into account, the optimum value of the sensing time has
been found numerically. When a PU arrives, the SU must leave the
spectrum and continues its transmission on a free spectrum after
possibly some HOs. Clearly, multiple spectrum HOs will increase
the overall sensing time \cite{yuan2010}. In \cite{lee} it is
assumed that the licensed spectrums are numbered sequentially, and
the SU starts to sense the spectrums from top of a list. In the
case of occupation, the SU senses the next one, and this process
is continued until an idle spectrum is found. Then, in \cite{lee},
an optimization problem is formulated in order to minimize the
average sensing time. Although the false detection and spectrum
handover effects on sensing time have been investigated in
\cite{lee}, but the negative effect of the handover (equivalently
the effect of multiple sensing time) on the SU throughput has not
taken into account.

In this paper, we consider the energy detection (ED) method as PU
detection scheme and try to set appropriate values for the ED's
parameters, i.e., sensing time $\tau$ and decision threshold
$\lambda$. The same problem is formulated in \cite{liang} without
considering the impact of HO on the derived average throughput. In
fact, \cite{liang} assumes that the SU senses the one spectrum in
each time-slot and transmits on it if it is sensed free. As
mentioned before, there exist a trade-off on selecting a value for
the sensing time, and thus an optimization problem can be
formulated in order to choose an appropriate value for sensing
time. It is shown in \cite{liang} that the two dimensional
optimization problem (with respect to $\tau$ and $\lambda$) can be
simplified to a one-dimensional with respect to $\tau$. However,
in contrast to \cite{liang}, as we are considering the spectrum
mobility effect on the overall sensing time, we show that our
problem cannot convert to a one-dimensional one. In this paper, we
consider the sequential method for handover (SMHO) first
introduced in \cite{lee}. We evaluate the average sensing time,
the average number of required handover for a given maximum false
alarm probability, and the average throughput of a SU temporarily
used the spectrum allocated to $N_p$ primary users. We formulate
an optimization problem in which the optimal sensing time for
maximizing the SU throughput is obtained. Different from
\cite{liang}, we show that our problem cannot convert to a
one-dimensional optimization problem. Then, we propose a weighted
based scheme for handover (WBHO) as a trade-off between the
complexity of finding an optimal handover sequence and the maximum
achievable throughput of the SU. In the WBHO scheme considered, a
weight is assigned to each primary channel based on the channel
conditions and the PUs entrance probability in the next slots.
Then, the algorithm decreasingly sorts these channels based on
their weights. The WBHO scheme provides a higher average
throughput and lower consumed energy to find a transmission
opportunity compared to the SMHO.

The rest of this paper is organized as follows. Section II
describes the system model. Problem formulation and performance
analysis of the SMHO scheme are provided in Section III. In Section IV,
channel imperfections such as fading is considered, and we develop a
new weighted based handover framework. Numerical results
are presented in Section V, and finally the paper concludes in
Section VI.

\section{System Model}
We consider one secondary user and $N_{p}$ primary users, and in
each time-slot the SU user transmits on at most one of $N_{p}$
existing bands by using opportunistic methods. We assume the SU
always has packets to transmit, and therefore it will start
transmission when an opportunity is found. A thoroughly
synchronous system is assumed in this paper in which the SU is
synchronous in time-slots with the PUs. When a PU has no data for
transmission it does not use it�s time-slots, thus provides a
transmission opportunity for the SU. But if the PU has data for
transmission, it starts transmitting at the beginning of the next
time-slot. In order to find the transmission opportunities
appropriately and protecting the PUs from harmful interference,
the sensing process must be performed at the beginning of each
time-slot. We assume that the SU is equipped with a simple
transceiver, so they are able to sense only one channel per
time-slot. We also assume that there is a fixed time $\tau_{ho}$
for the SU detector to change its channel and switch to a new one
independent of the channel frequency in which it switches. We
assume that the different PUs activities are independent. The
state of channel $i$ that used by $i$-th PU at time-slot t is
denoted by $s_{i}(t)$:
\begin{equation}\label{eq1}
{s_i}\left( t \right) = \left\{{\begin{array}{*{20}{c}}
   {0{\rm{~}}:{\rm{~~if~channel~i~is~occupied~}}} \hfill  \\
   {1{\rm{~}}:{\rm{~~if~channel~i~is~idle~}}} \hfill  \\
\end{array}} \right.
\end{equation}

\begin{figure}[h]
  \begin{center}
        \includegraphics[width = 5 in]{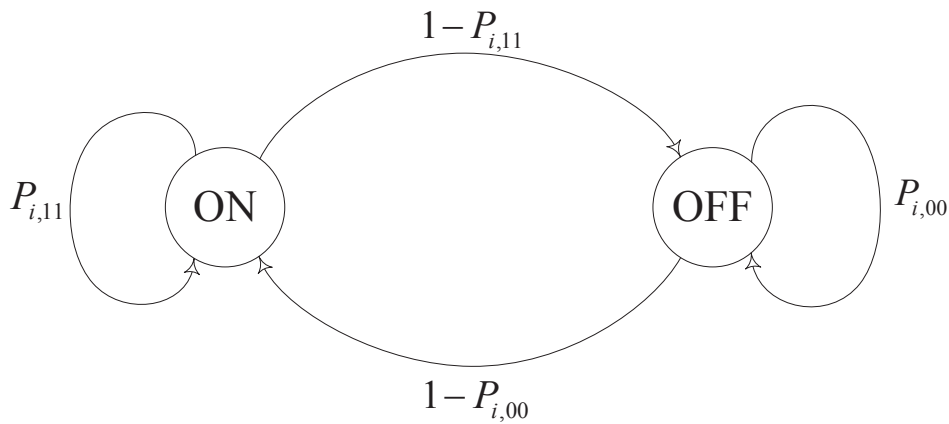}
        \caption{Graphical representation of ON-OFF primary user traffic model}
        \label{figonof}
        \end{center}
\end{figure}

The two state Markov model called ON-OFF traffic model, as shown
in Fig.~\ref{figonof}, can be used to model the correlation of a
channel states \cite{neely09}, where ON and OFF states in
Fig.~\ref{figonof} represent the presence and the absence of the
$i$-th PU, i.e., $s_i=1$ and $s_i=0$, respectively. $P_{i,00}$ and
$P_{i,11}$ are the probabilities that the channel state transits
from idle (in the current slot) to idle (in the next slot) and
from busy to busy, respectively. The transition probabilities of
the PUs in the ON-OFF model can be determined according to traffic
statistics from the long-term observation. Note that for $i$-th
PU, the steady state idle probability of the channel $i$ can be
calculated as:
\begin{equation}\label{eqn2}
\Pr \left\{ {{s_i}\left( t \right) = 0} \right\} = {P_{i,0}} =
\frac{{1 - {P_{i,00}}}}{{2 - {P_{i,00}} - {P_{i,11}}}}
\end{equation}

Spectrum sensing can be formulated as a binary hypothesis testing
problem \cite{liang},
\begin{equation}\label{eq2}
\left\{ {\begin{array}{*{20}{c}}
   {{\mathcal{H}_0~:~y(n)=z(n)}{\rm{~:~~channel~is~idle}}} \hfill  \\
   {{\mathcal{H}_1~:~y(n)=u(n)+z(n)}{\rm{~:~~channel~is~occupied }}} \hfill  \\
\end{array}} \right.
\end{equation}
where the noise $z(n)$ is zero mean complex-valued, independent
and identically distributed (i.i.d) Gaussian sequence, $u(n)$ is
the PUs signal and independent of $z(n)$, and $y(n)$ is the $n$-th
sample of the received signal. Generally, there are various PU
detection schemes such as match filtering, cyclostationary feature
detection, waveform-based sensing, and energy detection (ED)
\cite{arsalan09},

among which the ED is the most prevalent because of its low
complexity and ease of implementation. Further, it does not
require any information about the PUs' signal attributes
\cite{urkowitz}. By defining $X$ as a decision metric for the ED,
we have,
\begin{equation}\label{eq3}
{X} = \sum\limits_{n = 1}^N {{{\left| {y\left( n \right)} \right|}^2}}
\end{equation}
where $X$ is the received energy in the detector. $N$ represents
the number of samples that is equal to $N = \tau {f_s}$, where
$\tau$ and $f_s$ are the sensing time and the sampling frequency,
respectively. Finally, the decision criteria is defined as,
\begin{equation}\label{eq4}
\left\{ {\begin{array}{*{20}{c}}
   {X < \lambda {\rm{  }} \equiv {\mathcal{H}_0}} \hfill  \\
   {X \ge \lambda {\rm{  }} \equiv {\mathcal{H}_1}} \hfill  \\
\end{array}} \right.
\end{equation}

Let $\sigma _u^2$ denote the received energy of the PU signal, and
$\sigma _{z}^{2}$ represent the noise variance, then the received
signal to noise ratio due to the PU is computed as $\gamma  =
\frac{{\sigma _u^2}}{{\sigma _z^2}}$. Assume that $\lambda$ is the
threshold of the ED decision rule, and $\overline {P_d} $ is the
minimum allowable probability of detection. If the number of
received signal's samples is large enough, the statistical
distribution of $X$ can be approximated by a Gaussian
distribution, and then the detection and false alarm probabilities
can be computed using the following formulas \cite{liang}:
\begin{equation}\label{eq5}
{P_d} = Q\left[ {\left( {\frac{\lambda }{{\sigma _u^2}} - 1 -
\gamma } \right)\sqrt {\frac{{\tau {f_s}}}{{1 + 2\gamma }}} }
\right]
\end{equation}
\begin{equation}\label{eq6}
{P_{fa}} = Q\left[ {\left( {\frac{\lambda }{{\sigma _u^2}} - 1}
\right)\sqrt {\tau {f_s}} } \right],
\end{equation}
where $P_d$ and $P_{fa}$ are the detection and false alarm
probabilities, respectively. For ${P_d} = \overline {{P_d}} $,
which $\overline {{P_d}} $ is a fixed amount, we obtain:
\begin{equation}\label{eq7}
\lambda  = \left( {{Q^{ - 1}}\left( {\overline {{P_d}} }
\right)\sqrt {\frac{{1 + 2\gamma }}{{\tau {f_s}}}}  + 1 + \gamma }
\right)\sigma _u^2
\end{equation}

Suppose that $\beta  = {Q^{ - 1}}\left( {\overline {{P_d}} }
\right)\sqrt {1 + 2\gamma } $, (\ref{eq7}) is simplified to:
\begin{equation}\label{eq8}
{P_{fa}} = Q\left[ {\beta  + \gamma \sqrt {\tau {f_s}} } \right]
\end{equation}

For the SU, each slot contains two phases: 1) sensing phase, and
2) transmission phase. The sensing phase contains several
mini-slots of duration $\tau$ (sensing time of each channel).
Sensing is carried out by the SU in mini-slots, and once the
transmission opportunity is found, the transmission phase will be
started. This kind of access, i.e., listen-before-talk (LBT) is a
common method in many wireless communication systems; see e.g. the
quiet period in \emph{IEEE 802.22} standard \cite{stevenson}. The
sensing procedure is performed in an order based on the predefined
sensing sequence. Given the primary-free probabilities, i.e.,
${P_{0,j}}{\rm{~}},{\rm{~}}1 \le j \le {N_p}$, in this paper we
aim to formulate the performance of the SU for two kinds of the
sensing sequence and find the optimal setting values for the ED

\section{Performance Analysis}

In this section, we first consider the SMHO scheme and derive its
performance, and then the WBHO scheme is addressed. As mentioned
above, in the SMHO scheme, the SU arranges the frequency channels
by their numbers. If HO is required, the SU must sense the
spectrum from the top of the list and if it is sensed free, the SU
begins to transmit on it. In other case, the SU senses the next
spectrum, and this scenario continues until an idle spectrum is
found.

First we compute the average number of HOs, denoted by $\overline
g _{SMHO}$, to find an idle spectrum. The SU transmits on the
$k$-th channel with the following probability,
\begin{equation}\label{eq9}
\begin{array}{c}
 {q_k} = \Pr \left\{ {{\rm{ED~says~}}1|{\mathcal{H}_0}} \right\}{P_{k,0}} + \Pr \left\{ {{\rm{ED~says~}}1|{\mathcal{H}_1}} \right\}{P_{k,1}} \\
  = {P_{fa}}{P_{k,0}} + {P_d}{P_{k,1}}, \\
 \end{array}
\end{equation}
where ${P_{k,0}}$ and ${P_{k,1}}$ are the absence and presence
probability of the $k$-th PU, respectively.

\emph{Lemma 1}: The average number of HOs is equal to
\begin{equation}\label{eq10}
{\overline g _{SMHO}}  =  {q_1}{\left( 1 - q_2 \right)} +
2{q_1}{q_2}{\left( 1 - q_3 \right)} +  \cdots  + \left( \alpha - 1
\right) \left( 1 - q_{\alpha} \right) \prod\limits_{j = 1}^{\alpha
-1}  {q_j} + \alpha \prod\limits_{j = 1}^\alpha  {{q_j}}
\end{equation}
where $\alpha$ is the maximum number of allowable HOs.

\emph{Proof}: According to the SMHO, if the SU transmits on the
$i$-th spectrum, $\left( i - 1 \right)$ times hand over will be
necessary. The probability of $\left( i - 1 \right) $ consecutive
channel are sensed busy, and the $i$-th channel is sensed free is
equal to $\left( 1 - q_{i} \right) \prod\limits_{k = 1}^{i-1}
{{q_k}}$. In addition, there are two constraints on the maximum
number of HOs, the first one is due to the number of channels (or
equivalently the number of PUs), where the number of sensed
channels cannot exceed the number of the PUs, and the second is
due to the time-slot period, where the sum of the elapsed times
for both sensing and HO procedures can not exceed the time-slot
duration. So, we have,
\begin{equation}\label{eq11}
\alpha  = \min \left( {\left\lfloor {\frac{{T - \tau }}{{\tau  + {\tau _{ho}}}}} \right\rfloor ,{N_P} - 1} \right){\rm{ }}
\end{equation}
where $\tau_{ho}$ is defined above and $T$ is the duration of each
time-slot. Thus, the average number of HOs can be determined by
(\ref{eq10}).

Then, we can easily conclude the following lemma.

\emph{Lemma 2}: The average time of spectrum sensing can be
calculated as, $E\left\{{\rm{sensing~time}} \right\} = \tau  +
{\overline g _{SMHO}}\left( {\tau  + {\tau _{ho}}} \right)$.

In the following Lemma, the average achievable throughput has been
given.

\emph{Lemma 3}: Considering the maximum allowable number of HOs
equal to $\alpha$, the average achievable normalized throughput
can be calculated as,
\begin{equation}\label{eq12}
R = \sum\limits_{m = 0}^\alpha  {\underbrace {\left( {{C_1}{P_{m +
1,1}}\left( {1 - {P_d}} \right) + {C_0}{P_{m + 1,0}}\left( {1 -
{P_{fa}}} \right)} \right)}_{T1}\underbrace {{q_0}{q_1} \cdots
{q_m}\left( {1 - \frac{{\tau  + m\left( {\tau  + {\tau _{ho}}}
\right)}}{T}} \right)}_{T2}}
\end{equation}
where ${q_0} \buildrel \Delta \over = 1$, ${C_0} = {\log _2}\left(
{1 + {\gamma _s}} \right)$ and ${C_1} = {\log _2} \left({1 +
\frac{{{P_s}}}{{{N_0} + {P_p}}}} \right) = {\log _2} \left({1 +
\frac{{{\gamma _s}}}{{1 + {\gamma _p}}}}\right) $ are the SU's
capacity under the hypothesis $\mathcal{H}_0$ and $\mathcal{H}_1$,
respectively. ${\gamma _s}$ and ${\gamma _p}$ are the received
SNRs due to the secondary and primary user's signals at the SU
receiver, respectively, and $P_{k,0}$ is defined in (\ref{eqn2}).

\emph{Proof}: The proof is given in appendix A.

The optimum throughput can be obtained by solving the following
optimization problem $P1$:
\begin{equation}\label{eq13}
\begin{array}{l}
 P1:{\rm{~~}}\mathop {\max }\limits_{\tau ,\lambda } {\rm{~~~~}} R \\
 {\rm{~~~~~~~~~~}}s.t.{\rm{~}}\left\{ {\begin{array}{*{20}{c}}
   {{P_{fa}} \le P_{fa}^{\max }}  \\
   {{P_d} \ge P_d^{\min }}  \\
   {0 < \tau  < T}  \\
\end{array}} \right. \\
 \end{array}
\end{equation}

In \cite{liang}, the same problem is formulated without
considering the impact of HO on the throughput. In fact,
\cite{liang} assumes that the SU senses one spectrum in each
time-slot and transmits on it if it is detected free. As mentioned
before, the sensing accuracy, i.e., $P_d$ and $\left( {1 -
{P_{fa}}} \right) $, increases when $\tau$ increases. With the
increment of $\tau$, the time remained in each time-slot for
transmission reduces, which can lead to the throughput reduction.
As a consequence, the throughput decreases. Therefore setting an
appropriate value for sensing time used by the ED scheme is
necessary. The authors claim that the optimal value of $\lambda$
can be obtained by the maximum acceptable level of the false alarm
probability, and here by their problem simplifies to a
one-dimensional optimization problem. In the following we show
that our optimization problem cannot convert to a one-dimensional
one.

Assume that ${\lambda _1} < {\lambda _0}$, so based on (\ref{eq5})
and (\ref{eq6}), ${P_d}\left( {\tau ,{\lambda _1}} \right) >
{P_d}\left( {\tau ,{\lambda _0}} \right)$, and ${P_{fa}}\left(
{\tau ,{\lambda _1}} \right) > {P_{fa}}\left( {\tau ,{\lambda _0}}
\right)$. Then, from (\ref{eq9}), we have,
\begin{equation}\label{neweq1}
q\left( {\tau ,{\lambda _1}} \right) > q\left( {\tau ,{\lambda _0}} \right)
\end{equation}

Therefore, we can conclude that if $\lambda$ decreases, $q$
increases, but the term $\left( {1 - \frac{{\tau  + m\left( {\tau
+ {\tau _{ho}}} \right)}}{T}} \right)$ does not change, and
consequently $T2$ which is defined in (\ref{eq12}) increases as
well. On the other hand, the term $T1$ defined in (\ref{eq12})
decreases, as $\lambda$ decreases. Considering the above
conflicting effects, we must choose an appropriate value for
$\lambda$ based on the constraints of the $P1$. In the following,
we convert our two-dimensional optimization problem to a
one-dimensional one by using an acceptable value for detection
probability.

Supposing ${P_d} = P_d^{\min }$, the optimization problem convert
to:
\begin{equation}
\begin{array}{l}
 P2:{\rm{~~}}\mathop {\max }\limits_{\tau ,\lambda } {\rm{~~~~}} R \\
 {\rm{~~~~~~~~~~}}s.t.{\rm{~}}\left\{ {\begin{array}{*{20}{c}}
   {{P_{fa}} \le P_{fa}^{\max }}  \\
   {{P_d} = P_d^{\min }}  \\
   {0 < \tau  < T}  \\
\end{array}} \right. \\
 \end{array}
\end{equation}

It is worth noting that, from (\ref{eq7}) and (\ref{eq8}), under
the assumption ${P_d} = P_d^{\min }$, the throughput of the SU
derived in (\ref{neweq12}) only depends on $\tau$.

In order to satisfy the first constraint from (\ref{eq8}), we must have,
\begin{equation}
Q\left[ {\beta  + \gamma \sqrt {\tau {f_s}} } \right] \leqslant P_{fa}^{\max }{\text{~}}
\end{equation}
so
\begin{equation}\label{taumin}
\tau  \geqslant \frac{1}{{{f_s}}}{\left( {\frac{{{Q^{ -
1}}\left({P_{fa}^{\max }} \right) - \beta }}{\gamma }} \right)^2}
\end{equation}
where $\beta  = {Q^{ - 1}}\left( {P_{d}^{\min }} \right)\sqrt {1 +
2\gamma } $, and $\frac{1}{{{f_s}}}{\left( {\frac{{{Q^{ -
1}}\left( {P_{fa}^{\max }} \right) - \beta }}{\gamma }}
\right)^2}$ can be considered as ${\tau _{\min }}$. Therefore, the
problem $P2$ can be easily simplified as $P3$,
\begin{equation}\label{optimprob}
\begin{array}{l}
  P3:{\rm{~~}}\mathop {\max }\limits_\tau  {\rm{~~~~}} R \\
 {\rm{~~~~~~~~~~}}s.t.{\rm{~~~}}{\tau _{\min }} < \tau  < T \\
 \end{array}
\end{equation}
where ${\tau _{\min }} = \frac{1}{{{f_s}}}{\left( {\frac{{{Q^{ -
1}}\left( {P_{fa}^{\max }} \right) - \beta }}{\gamma }}
\right)^2}$.

\emph{Proposition}: The SU's maximum achievable throughput is
saturated by the increment of the number of primary users.

\emph{Proof}: In the optimization problem described by
(\ref{eq12}), (\ref{taumin}), and (\ref{optimprob}), the number of
PUs only manifests itself on $\alpha$. For a small value of $N_p$,
increasing the number of PUs leads to the increment of $\alpha$,
and consequently the increase in the SU's maximum achievable
throughput. Considering the constraint of derived optimization
problem imposed by sensing time, i.e, $\tau _{min} < \tau < T$,
for $N_p \geq \left\lfloor {\frac{{T - \tau _{min} }}{{\tau _{min}
+ {\tau _{ho}}}}} \right\rfloor + 1$, $ \alpha = \left\lfloor
{\frac{{T - \tau }}{{\tau  + {\tau _{ho}}}}} \right\rfloor $,
where it will be independent of $N_p$. Therefore, for such values
of $N_p$, the maximum achievable throughput does not improve and
will be saturated.



\section{Impact of Channel Fading}

In the previous section, a HO scheme for AWGN channel is introduced.
Considering channel imperfections, we aim to modify the derived optimization problem, and then develop a new systematic channel weighting algorithm to create an appropriate sensing sequence, which provides an average throughput for the SU higher than the SMHO throughput.

To extend the SMHO scheme for the case of presence of multipath fading, (\ref{eq12}) is modified to:
\begin{equation}\label{neweq12}
\overline R  = \iint\limits_{{\gamma _p},{\gamma _s}} {R{\text{
}}{f_{{\gamma _p},{\gamma _s}}}\left( {{\gamma _p},{\gamma _s}}
\right)d{\gamma _p}d{\gamma _s}}
\end{equation}

Therefore, the optimal throughput can be obtained by solving the following
optimization problem:
\begin{equation}
\begin{array}{l}
 \mathop {\max }\limits_\tau  {\rm{~~~~}} \overline R \\
 {\rm{~}}s.t.{\rm{~~~}}{\tau _{\min }} < \tau  < T \\
 \end{array}
\end{equation}

However, the main disadvantage of the SMHO scheme is its channel searching strategy, which regardless of the quality of the $j$-th channel, it cannot be sensed by the SU until all $\left( j - 1 \right)$ previous channels has been sensed. To develop a more appropriate and practical handover framework, we assume that the SU equipped by a transceiver with adaptive modulation and coding (AMC) capability. In the first step, we model each of the PU's traffic and channel with two state ON-OFF and the $K$-state Markov process, respectively. To incorporate a proper HO framework, we must address three
questions in our proposed scheme. First, when the SU must vacate
its currently used channel? Second question is that which channel
should be sensed at first? And finally, how much computational
burden is imposed by the HO scheme on the SU?

To find a general solution to cover the first question, we define
a factor named ${E_i}\left( p \right) = {s}$. That is, the $i$-th
PU will arrive in the $s$ following time-slots with the
probability of $p$. Suppose that the SU would be able to predict
the probability of the $i$-th PUs entrance in the next time-slots
by using some PU traffic prediction algorithms \cite{yuan2010},
\cite{neely09}. In this case, for predefined values of $s _0$ and
$p_0$, the SU calculates $p$ in which ${E_i}\left( P \right) =
{s_0}$, and the HO procedure is started if $ p > p_0 $. $s_0$ and
$p_0$ are design parameters depending on the required QoS for the
SU, the maximum level of interference to the PUs, and the average
time required for the HO procedures by which the SU finds a new
transmission opportunity and resumes its unfinished transmission.

\begin{figure}
  \begin{center}
        \includegraphics[width = 6.5 in]{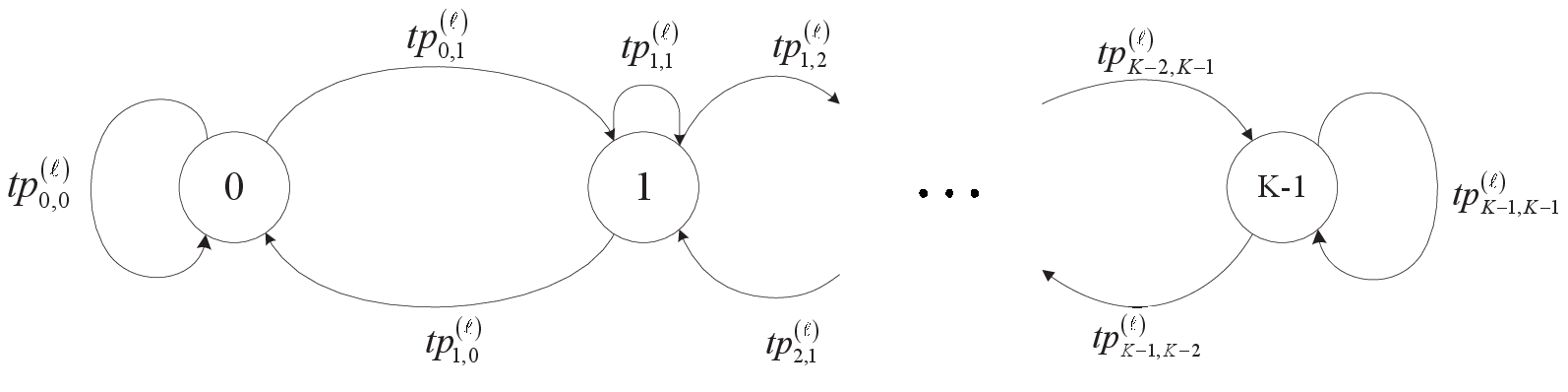}
        \caption{Illustration of $K$-state Markov chain}
        \label{channel_model}
        \end{center}
\end{figure}

In the ON-OFF model which is exploited to model the PUs activity
in this paper, the presence and absence probabilities of the PUs
will be obtained by long term observation and will not
significantly change in short term. Let ${S^{\left( \ell  \right)}} = \left\{ {s_1^{\left( \ell
\right)},s_2^{\left( \ell  \right)}, \ldots ,s_K^{\left( \ell
\right)}} \right\}$ denote a set of $K$ states of the $\ell $-th
primary channel and $\left\{ {S_t^\ell } \right\},{\rm{ }}t = 0,1,
\ldots $ be a constant Markov process, which has stationary
transitions \cite{wang95}. Assume that $\pi _i^{\left( \ell
\right)}$ and $tp_{ij}^{\left( \ell \right)}$ represent the
steady-state probability of $i$-th state and the state transition
probability from the $i$-th state to the $j$-th state of the
$\ell$-th channel. For all $i,j \in \left\{ {0,1, \ldots ,K - 1}
\right\}$, we have,
\begin{equation}\label{neq19}
\pi _i^{\left( \ell  \right)} = \Pr \left\{ {S_t^{\left( \ell
\right)} = s_i^{\left( \ell  \right)}} \right\}
\end{equation}
and
\begin{equation}\label{neq20}
tp_{ij}^{\left( \ell  \right)} = \left\{ {\begin{array}{*{20}{c}}
   {\Pr \left\{ {S_{t + 1}^{\left( \ell  \right)} = s_j^{\left( \ell  \right)}|S_t^{\left( \ell  \right)} = s_i^{\left( \ell  \right)}} \right\},} \hfill & {for{\rm{~}}\left| {i - j} \right| \le 1} \hfill  \\
   {0{\rm{~}},} \hfill & {O.W.} \hfill  \\
\end{array}} \right.
\end{equation}
where $\sum\limits_{j = 0}^{K - 1} {tp_{ij}^{\left( \ell \right)}}
= 1$ and $\sum\limits_{i = 0}^{K - 1} {\pi _i^{\left( \ell
\right)}}  = 1$. Fig.~\ref{channel_model} shows the assumed
$K$-state Markov chain for the $\ell$-th channel. Rayleigh
distribution is a conventional model for the received signal
envelop in a typical multipath propagation channel. It can be
shown that the received SU's SNR is proportional to the square of
the signal envelop and exponentially distributed with the
following probability density function \cite{wang95} and
\cite{zhang99},
\begin{equation}\label{neq21}
p\left( {{\gamma _s}} \right) = \frac{1}{{\overline {{\gamma _s}}
}}\exp \left( { - \frac{{{\gamma _s}}}{{\overline {{\gamma _s}}
}}} \right)
\end{equation}
where $\overline {{\gamma _s}} $ is both the mean and standard
deviation of the SU's SNR. Let $0 = \gamma _{s,\ell }^{\left( 0
\right)} < \gamma _{s,\ell }^{\left( 1 \right)} <  \cdots  <
\gamma _{s,\ell }^{\left( {K - 1} \right)} = \infty $ be the
quantized SNR levels for the $\ell$-th channel. The channel will
be in the state ${s_m^{\left( \ell  \right)}}$, if the received
SNR is placed within the interval of $\left[ {\left. {\gamma
_{s,\ell }^{\left( m \right)},\gamma _{s,\ell }^{\left( {m + 1}
\right)}} \right)} \right.$. Considering (\ref{neq21}) the
steady-state probability of each state can be computed as,
\begin{equation}\label{neq22}
\pi _m^{\left( \ell  \right)} = \int_{\gamma _{s,\ell }^{\left( m
\right)}}^{\gamma _{s,\ell }^{\left( {m + 1} \right)}}
{\frac{1}{{\overline {{\gamma _{s,\ell }}} }}\exp \left( { -
\frac{x}{{\overline {{\gamma _{s,\ell }}} }}} \right)dx}
\end{equation}

The transition probabilities can be calculated as \cite{wang95},
\cite{zhang99}
\begin{equation}\label{neq23}
tp_{m,m + 1}^{\left( \ell  \right)} \approx \frac{{N_{m +
1}^{\left( \ell  \right)}T}}{{\pi _m^{\left( \ell
\right)}}}{\rm{~}},{\rm{~~}}m = 0,1, \ldots ,K - 2
\end{equation}
and
\begin{equation}\label{neq24}
tp_{m,m - 1}^{\left( \ell  \right)} \approx \frac{{N_m^{\left(
\ell  \right)}T}}{{\pi _m^{\left( \ell
\right)}}}{\rm{~}},{\rm{~}}m = 1,2, \ldots ,K - 1
\end{equation}
where
\begin{equation}\label{neq25}
N_m^{\left( \ell  \right)} = \sqrt {\frac{{2\pi \gamma _{s,\ell
}^{\left( m \right)}}}{{\overline {{\gamma _{s,\ell }}} }}}
{f_d}\exp \left( { - \frac{{\gamma _{s,\ell }^{\left( m
\right)}}}{{\overline {{\gamma _{s,\ell }}} }}} \right)
\end{equation}
where $N_m$ denotes the level crossing rate, and $f_d$ represents
the maximum Doppler frequency, which can be calculated by knowing
the moving speed of the mobile terminal, the speed of the light,
and the carrier frequency. Other transition probabilities are
given by,
\begin{equation}\label{neq26}
tp_{m,m}^{\left( \ell  \right)} = \left\{ {\begin{array}{*{20}{c}}
   {1 - tp_{m,m + 1}^{\left( \ell  \right)}{\rm{~}},} \hfill & {if{\rm{~}}m = 0} \hfill  \\
   {1 - tp_{m,m - 1}^{\left( \ell  \right)}{\rm{~}},} \hfill & {if{\rm{~}}m = K - 1} \hfill  \\
   {1 - tp_{m,m - 1}^{\left( \ell  \right)} - tp_{m,m + 1}^{\left( \ell  \right)}{\rm{~}},} \hfill & {O.W.} \hfill  \\
\end{array}} \right.
\end{equation}

The expected throughput in the next $s$ slots for the $ \ell $-th
channel can be computed as,
\begin{equation}\label{neq27}
\begin{array}{c}
 E\left\{ R \right\}_s^{\left( \ell  \right)} = \underbrace {\Pr \left\{ {S_1^{\left( \ell  \right)} = s_0^{\left( \ell  \right)}} \right\}r_0^{\left( \ell  \right)} + \Pr \left\{ {S_1^{\left( \ell  \right)} = s_1^{\left( \ell  \right)}} \right\}r_1^{\left( \ell  \right)} +  \cdots  + \Pr \left\{ {S_1^{\left( \ell  \right)} = s_{K-1}^{\left( \ell  \right)}} \right\}r_{K-1}^{\left( \ell  \right)}}_{{\rm{1 - th~next~time~slot}}} +  \\
 \underbrace {\Pr \left\{ {S_2^{\left( \ell  \right)} = s_0^{\left( \ell  \right)}} \right\}r_0^{\left( \ell  \right)} + \Pr \left\{ {S_2^{\left( \ell  \right)} = s_1^{\left( \ell  \right)}} \right\}r_1^{\left( \ell  \right)} +  \cdots  + \Pr \left\{ {S_2^{\left( \ell  \right)} = s_{K-1}^{\left( \ell  \right)}} \right\}r_{K-1}^{\left( \ell  \right)}}_{{\rm{2 - th~next~time~slot}}} +  \\
  \vdots  \\
 \underbrace {\Pr \left\{ {S_s^{\left( \ell  \right)} = s_0^{\left( \ell  \right)}} \right\}r_0^{\left( \ell  \right)} + \Pr \left\{ {S_s^{\left( \ell  \right)} = s_1^{\left( \ell  \right)}} \right\}r_1^{\left( \ell  \right)} +  \cdots  + \Pr \left\{ {S_s^{\left( \ell  \right)} = s_{K-1}^{\left( \ell  \right)}} \right\}r_{K-1}^{\left( \ell  \right)}}_{{\rm{s - th~next~time~slot}}} \\
 \end{array}
\end{equation}
where ${r_i^{\left( \ell  \right)}}$ denotes the transmission rate
in the $i$-th state of the $\ell$-th channel due to exploited
adaptive modulation coding scheme, and $\Pr \left\{ {S_j^{\left(
\ell  \right)} = s_i^{\left( \ell  \right)}} \right\} = \pi
_i^{\left( \ell  \right)}$ for all $1 \le j \le s{\rm{~}},0 \le i
\le K-1$. So we can simplify (\ref{neq27}) as,
\begin{equation}\label{neq28}
E\left\{ R \right\}_s^{\left( \ell  \right)} = s\left( {\pi
_0^{\left( \ell  \right)}r_0^{\left( \ell  \right)} + \pi
_1^{\left( \ell  \right)}r_1^{\left( \ell  \right)} +  \cdots  +
\pi _{K-1}^{\left( \ell  \right)}r_{K-1}^{\left( \ell  \right)}}
\right)
\end{equation}

We define \emph{run} as a consecutive series of idle states
without occupied states. Hence, a \emph{run} is a period within a
SU could use the resource.
\begin{equation}\label{neq29}
\Pr \left[ {{\rm{run~length~}} \ge {\rm{~}}s} \right] = {\left( {1
- {P_{\ell , 00 }}} \right)^{s - 1}}
\end{equation}

Let ${AP_{\ell ,s}}$ be the absence probability of the $\ell$-th
PU in the next $s$ consecutive time-slot. Considering
(\ref{neq29}), we have,
\begin{equation}\label{neq30}
{AP_{\ell ,s}} = {P_{\ell , 0}} \times \Pr \left[
{{\rm{run~length~}} \ge {\rm{~}}s} \right] = {P_{\ell , 0 }}{\left(
{1 - {P_{\ell , 00 }}} \right)^{s - 1}}
\end{equation}

When the SU needs to perform the HO procedure, it assigns a weight
to each channel. Considering (\ref{neq28}) and (\ref{neq30}), the
SU assigns weight ${w_\ell }$  to the $\ell $-th channel using the
following formula:
\begin{equation}\label{neq31}
{w_\ell } = \frac{{E\left\{ R \right\}_s^{\left( \ell
\right)}}}{{{AP_{\ell ,s}}}}
\end{equation}

Then the SU sorts the channels based on their weights,
decreasingly and starts to sense the channels using the derived
sensing sequence. As a result, arranging spectrums by their
weights gives an opportunity to the SU to sense spectrums that are
more likely idle as well as higher expected throughput. Obviously,
we expect that the overall achievable throughput in the WBHO
scheme is higher than the SMHO approach.

\section{Simulation Results}
In this section, we first evaluate the performance of the SMHO
scheme by an exhaustive set of simulations to demonstrate the
effect of different parameters introduced throughout the paper. We
then consider the performance of the WBHO scheme and compare the
result with that of the SMHO scheme. To set up a simulation
environment, the values of SNR and sampling frequency are adopted
from \cite{liang}, $P_d^{\min }$ and $P_{fa}^{\max }$ are chosen
according to \emph{IEEE 802.22} \cite{stevenson}. These values are
given in Table.~\ref{table1}.

\begin{figure}
  \begin{center}
        \includegraphics[width = 6.5 in]{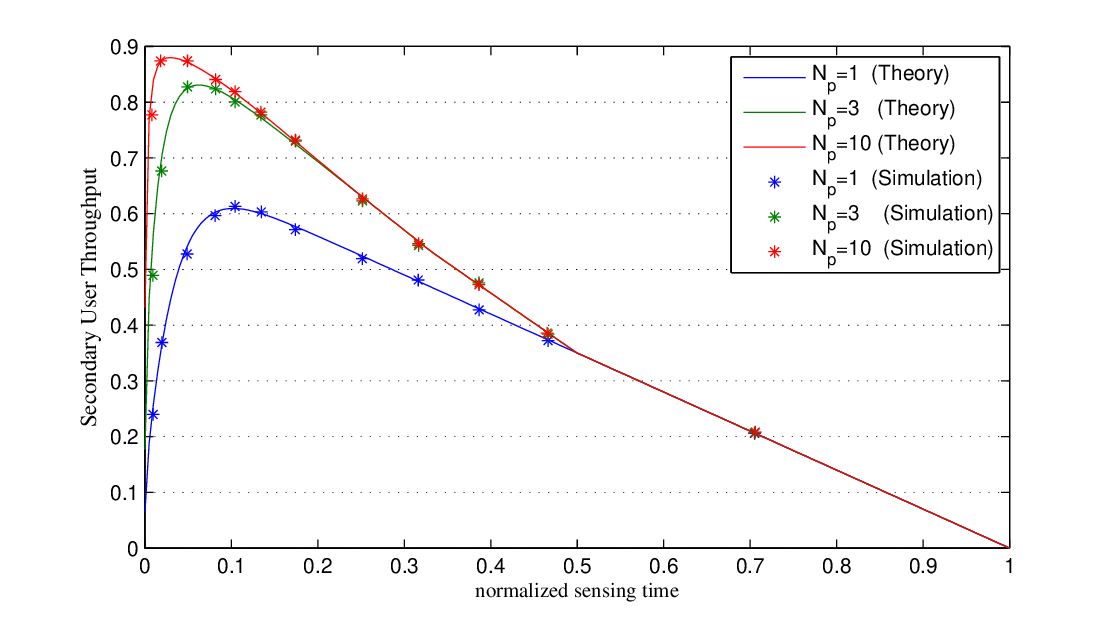}
        \caption{The average rate versus $\frac{\tau}{T}$ for various amount of $N_p$}
        \label{fig1}
        \end{center}
\end{figure}

The probabilities of the PUs' presence, i.e, ${P_{k,0}}~~{\rm{
}}\forall 1 \le k \le {N_p}$, are assumed to be identical and
equal to $0.65$. The average throughput has been computed after
simulating the scenario for $1200$ time-slots. Fig.~\ref{fig1}
verifies our analysis and illustrates the plots of achievable rate
of the SMHO versus the sensing time $\tau$ normalized to slot
period for different number of $N_p$ (the number of the PUs). For
a large sensing time $\tau$, the plots for different values of
$N_p$ coincide.

This behavior is expected due to our previous discussions on the
constraints which affect the number of possible HOs for a SU. As
stated in (\ref{eq11}), the number of possible HOs is dictated by
two factors; namely, the number of primary channels $N_p$, and the
ratio $\left( T - \tau \right) / \left( \tau + \tau _{ho}
\right)$. Therefore, as $\tau$ increases in Fig.~\ref{fig1}, we
observe that the second factor dominates and regardless of the
number of available primary channels $N_p$, the achieved
throughput becomes limited to a value corresponding to a lower
$N_p$. The observed coincidences of the plots in Fig.~\ref{fig1}
demonstrate this effect. The rate of the SU where there are $10$
primary channels equals the rate of the SU with $3$ primary
channels for approximately $\tau > 1/4 T$. Similarly, the rate
achieved by a SU with $3$ primary channels is equal to the rate of
a SU with only $1$ primary channel. Other important observations
can be made through Fig.~\ref{fig1}. First, since all the curves
posses a maxima, there exists an optimum value for the spectrum
sensing time. Second, as the number of primary channels increases,
the SU throughput increases as well, but in a saturating manner.

\begin{figure}
  \begin{center}
        \includegraphics[width = 6.5 in]{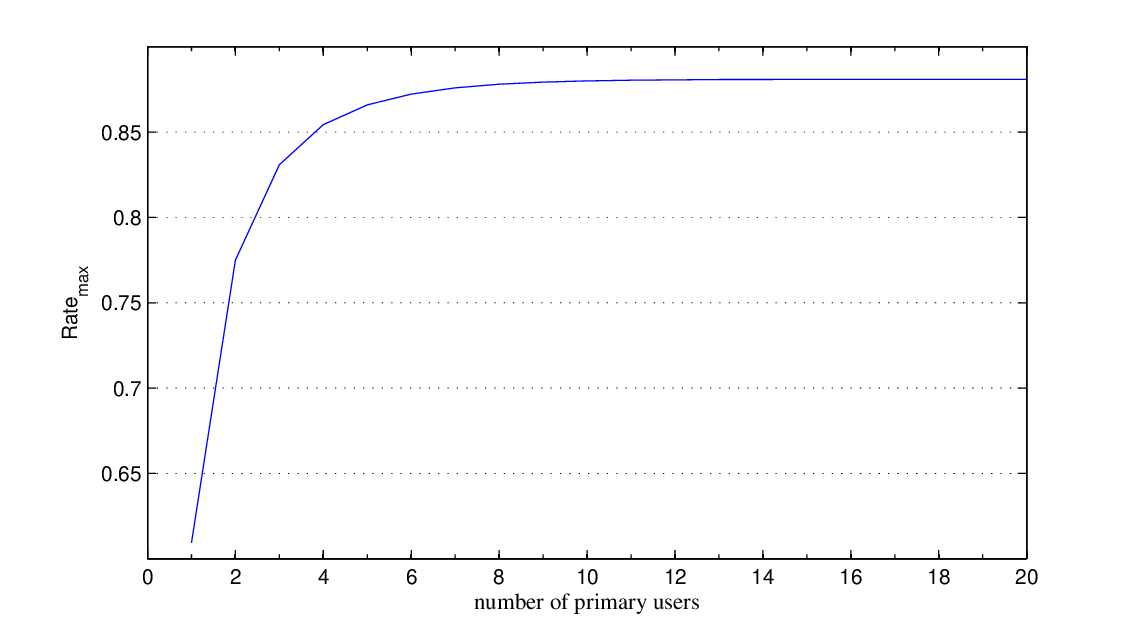}
        \caption{The maximum rate versus the number of primary users}
        \label{fig2}
        \end{center}
\end{figure}

Fig.~\ref{fig2} verifies the \emph{Proposition} and
demonstrates the plot of maximum possible achievable rate
(obtained by the optimum value of sensing time which is seen for
the $N_p = 1,~3,~10$ in the Fig.~\ref{fig1}) versus $N_p$. When
$N_p$ increases, the overall rate increases as well. On the other
hand, increasing in the value of $N_p$ leads to increment of the
average number of HOs, and as the number of HOs increases, the
transmission time reduces, so the maximum rate saturates.

\begin{figure}
  \begin{center}
        \includegraphics[width = 6.5 in]{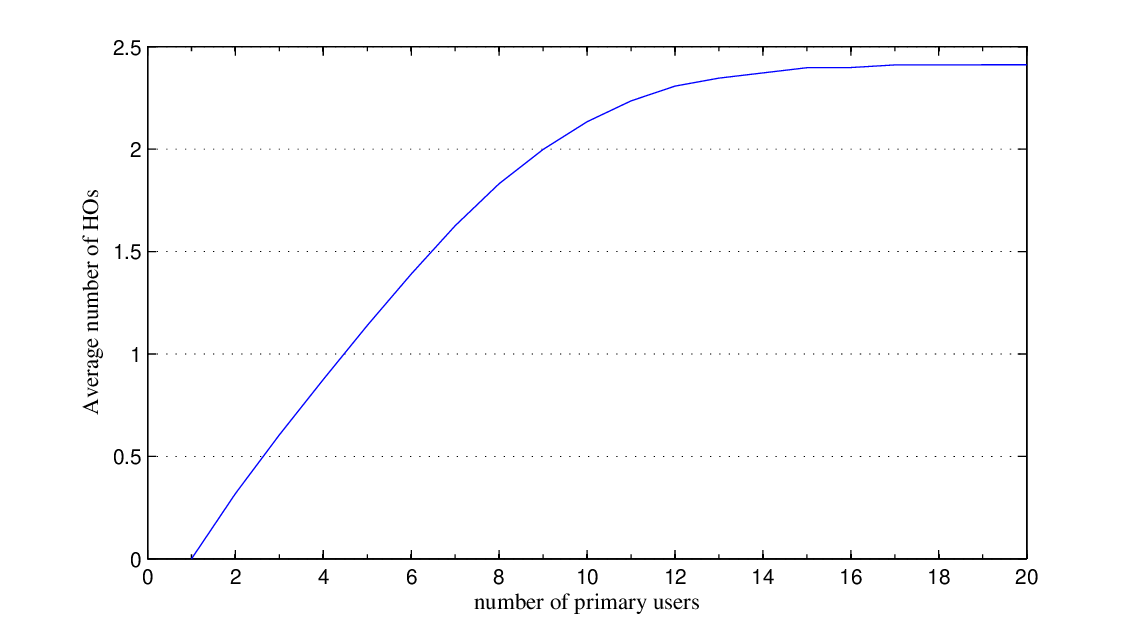}
        \caption{The average number of HOs versus the number of primary users}
        \label{fig3}
        \end{center}
\end{figure}

Fig.~\ref{fig3} illustrates the plot of average number of handover
versus $N_p$. From this Fig. we can see that as the number of
available channels increases, the average number of HOs in order
to find an idle channel, increases. That is, by the increment of
$N_p$, $\alpha$ in (\ref{eq11}) increases, and ${\overline g
_{SMHO}}$  in (\ref{eq10}) increases, consequently. However, as we
see in our further simulation, it does not lead to higher
throughput.

\begin{figure}
  \begin{center}
        \includegraphics[width = 6.5 in]{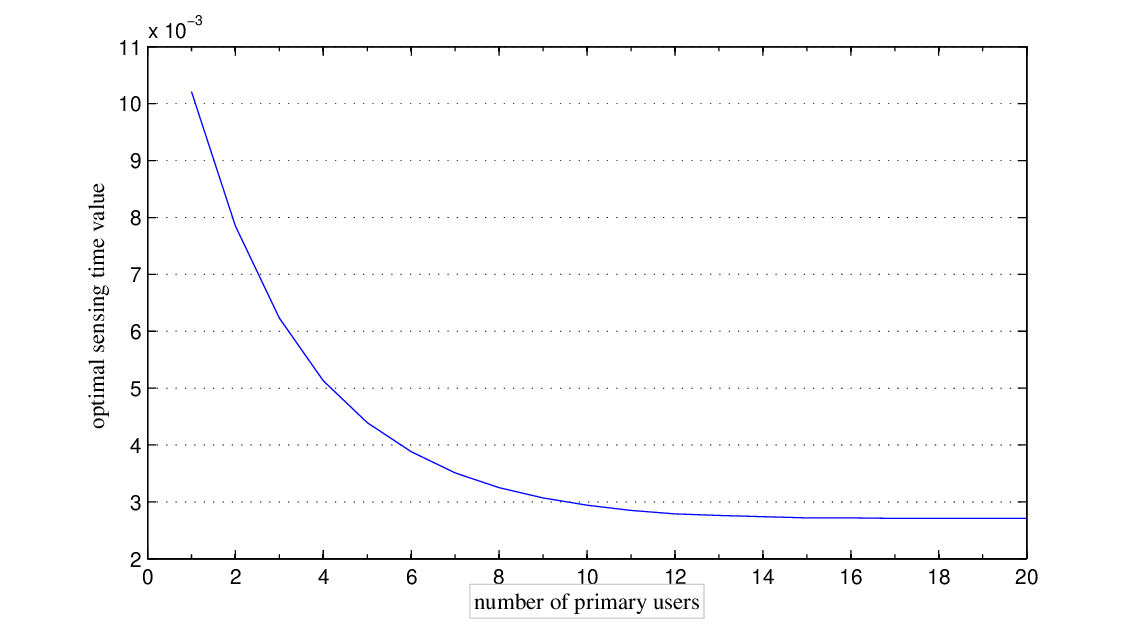}
        \caption{The optimum value for $\tau$ versus $N_p$}
        \label{fig4}
        \end{center}
\end{figure}

\begin{figure}
  \begin{center}
        \includegraphics[width = 6.5 in]{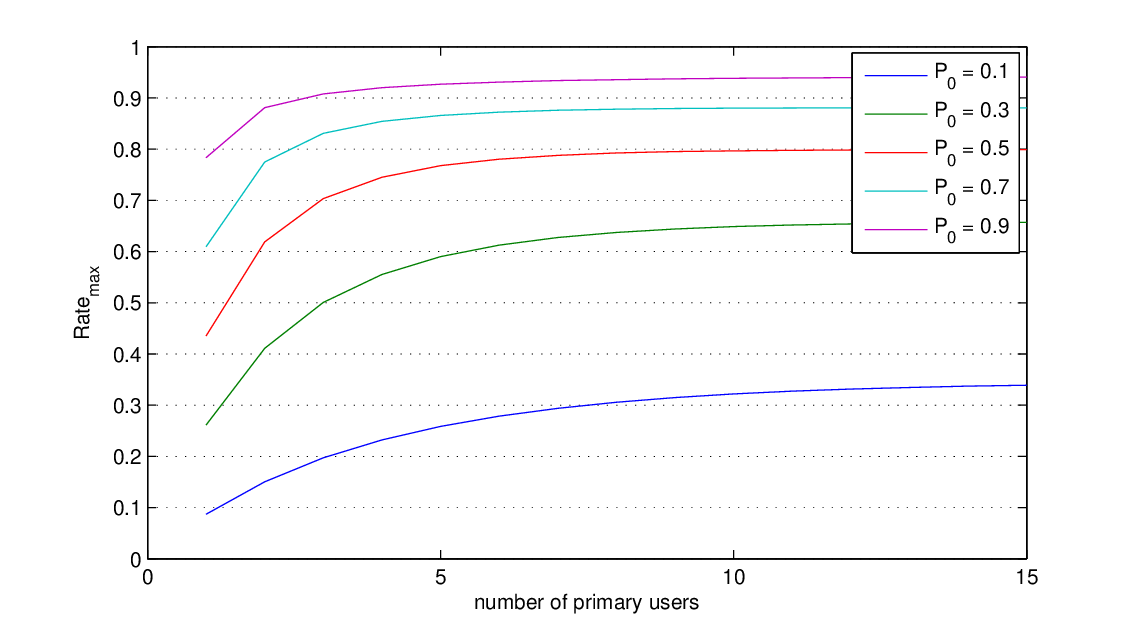}
        \caption{Maximum achievable rate versus $N_p$ for different values of $P_0$}
        \label{fig5}
        \end{center}
\end{figure}

Fig.~\ref{fig4} indicates that when $N_p$ increases the value of
$\tau_{opt}$, in which the maximum rate is achieved, decreases;
because the higher probability of finding idle spectrum is more
important than the sensing accuracy, as there are many spectrums
that the SU can utilize them with no priority. However, when
${N_P} \ge \left\lfloor {\frac{{T - {\tau _{\min }}}}{{{\tau
_{\min }} + {\tau _{ho}}}}} \right\rfloor  + 1$, by the increase
of $N_p$, the optimal value of sensing time, i.e., $\tau_{opt}$
and consequently the maximum achievable rate will not change, for
the same reason explained in the \emph{Proposition}. In
Fig.~\ref{fig5} the effect of PU's absence probability $(P_0)$ on
the achievable rate is shown. The increase of $P_0$ increases the
chance of finding a transmission opportunity and improves the SU's
throughput, as well.

In the following, we evaluate the performance of
the WBHO scheme and its advantages compared to the SMHO approach.


In order to simulate the WBHO scheme, the state transition
probability of each PU (in the ON-OFF model) is assumed to be a
uniform random variable within $0.1$ and $0.9$. Channel is
modelled via $11$-state Markov process using the same parameter as
Table~\ref{table1} in \cite{zhang99}. At the end of each time-slot
for the current channel, the SU calculates the $p$ in which $E_i
\left( p \right) = 3$ and compares it with $p_0 = \%90$. The HO
procedure is started provided that $p > p_0$. In this case, the SU
establishes a sorted set of the channels based on the weighted
computed in (\ref{neq31}) for $s = 5$ time-slot, and then starts
to sense the channels in order. For the SHMO scheme, the same
process has been performed, but the SU sorts channels based on
their numbers, sequentially. The average throughput is obtained
after $1200$ tim slots simulation.

\begin{figure}
  \begin{center}
        \includegraphics[width = 6.5 in]{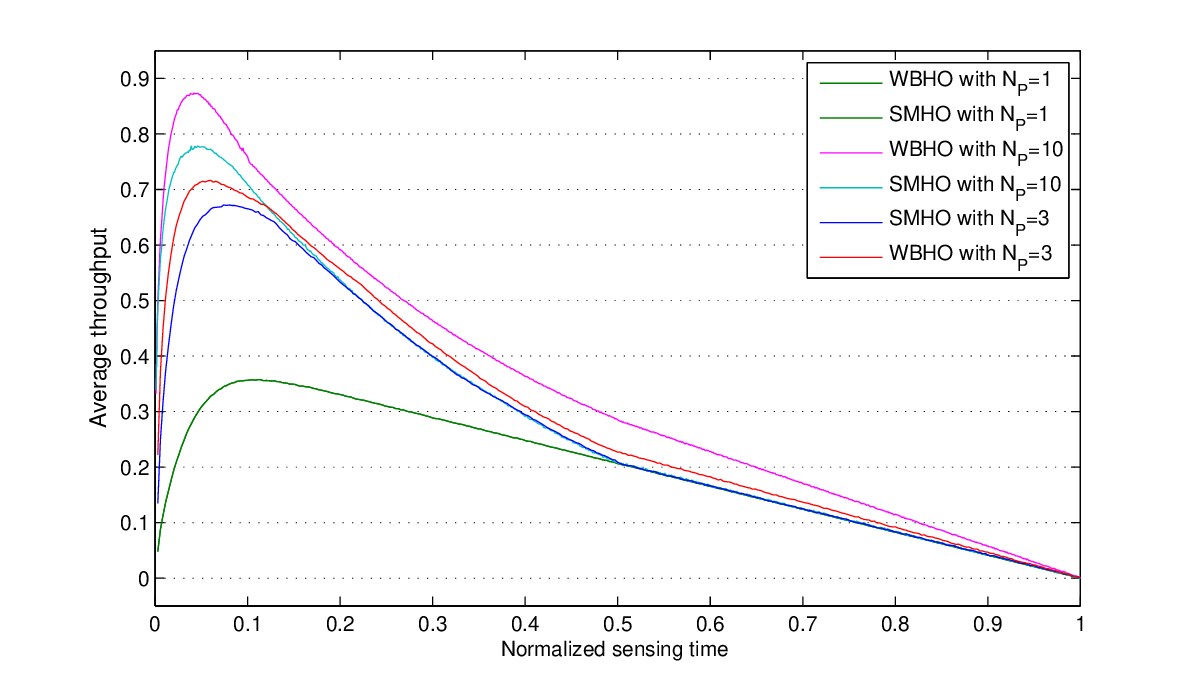}
        \caption{Comparison between the throughput of the SMHO and WBHO schemes for various values of $N_p$}
        \label{fig8}
        \end{center}
\end{figure}

\begin{figure}
  \begin{center}
        \includegraphics[width = 6.5 in]{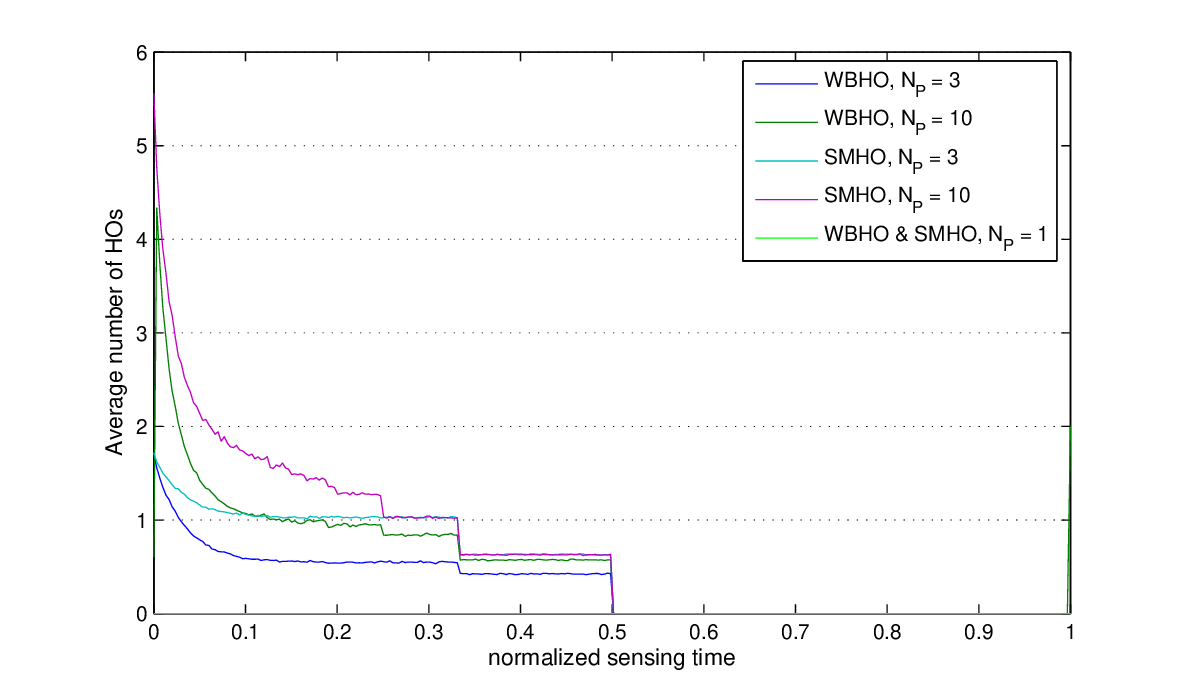}
        \caption{Comparison between the average number of HOs versus the sensing time for various number of $N_p$}
        \label{fig9}
        \end{center}
\end{figure}

Fig.~\ref{fig8} shows the plot of the SU's average achievable
throughput of the both schemes versus the sensing time. This
figure is noticeable in twofold: First, the results indicate the
better performance of the WBHO schemes. In fact, regardless of the
case $N_p = 1$, where two schemes offer the same throughput, the
SU can achieve higher average throughput by applying the WBHO
schemes for selecting its sensing sequence. Second, unlike the
SMHO, the throughput of the SU for different number of PUs is not
coincide; because having more PUs offers more spectrum bands with
different conditions which improves the chance of having a channel
with higher expected throughput, which depends on the PU absence
probability and the channel gain. While for $\tau > 0.5 T$,
regardless of the number of PUs, we cannot run the HO procedure
and sense more than one channel (see (\ref{eq11})), however, we
achieve a higher throughput by the increase of $N_p$ as a result
of having a higher probability to find a channel with a more
proper conditions. Finally, Fig.~\ref{fig9} represents the average
number of HOs versus sensing time for various number of the PUs.
Raising the number of the PUs leads to the increment of the number
of HOs required to find a transmission opportunity. Moreover,
average number of HOs reduces if the SU assigns more time to sense
a channel due constraint imposed by (\ref{eq10}). It is worth
mentioning that the average number of HOs for the WBHO scheme is
lower than the SMHO for all values of sensing time and the number
of the PUs. Therefore, the SU can achieve higher average
throughput with lower HOs and equivalently less consumed energy.

\section{Conclusion}
In this paper, we have considered the cognitive access of primary
channels by a secondary user. The average detection time by the
secondary user using SMHO and WBHO schemes have been evaluated. We have
formulated an optimization problem in order to find the optimum
sensing time in which the maximum throughput can be achieved. The
tradeoff between the maximum achievable throughput and the
consumed energy has been investigated. Finally, we have introduced
a design parameter to modify our optimization problem addressing
this tradeoff. Due to the new optimization problem, the acceptable
throughput can be achieved while the energy consumption is more
reasonable.

\appendix
Before proofing Lemma 3, we note that if the SU transmits on the
$(m+1)$-th channel (i.e., after $m$ times handover), the maximum
rate in the slot is calculated as
\begin{equation}\label{eq23}
 {r ^{\left( m \right)}} = {C_0}{P_{m+1,0}}\left( {1 - \frac{{ET_m}}{T}} \right)\left( {1 - {P_{fa}}\left( \tau  \right)} \right) +  {C_1}{P_{m+1,1}}\left( {1 - \frac{{ET_m}}{T}} \right)\left( {1 - {P_d}\left( \tau  \right)} \right)
\end{equation}
\begin{equation}\label{eq24}
ET_m = \tau  + m\left( {\tau  + {\tau _{ho}}} \right)
\end{equation}
where $ET_m$ is the time spent until the
SU chooses $(m+1)$-th spectrum for the transmission.

Now, we prove the lemma using the mathematical induction.
Let ${R^{\left( {k} \right)}}$ denote the
average normalized rate when the maximum number of allowable HOs is $k$.
We intend to show that $R^{\left( k \right)}$ can be calculated as
\begin{equation}\label{eq25}
R^{\left( k \right)} = \sum\limits_{m = 0}^k  {\left( {{C_1}{P_{m +
1,1}}\left( {1 - {P_d}} \right) + {C_0}{P_{m + 1,0}}\left( {1 -
{P_{fa}}} \right)} \right) {q_0}{q_1} \cdots
{q_m}\left( {1 - \frac{{\tau  + m\left( {\tau  + {\tau _{ho}}}
\right)}}{T}} \right)}
\end{equation}
For $k=0$, the maximum achievable rate can be calculated
as \cite{liang}
\begin{equation}
 {R^{(0)}} = {C_0}{P_0}\left( {1 - \frac{\tau }{T}} \right)\left( {1 - {P_{fa}}\left( \tau  \right)} \right) + {C_1}{P_1}\left( {1 - \frac{\tau }{T}} \right) \times  \left( {1 - {P_d}\left( \tau  \right)} \right)
\end{equation}

Suppose ${R^{\left( {k} \right)}}$ is true, we investigate the
validity of ${R^{\left( {k + 1} \right)}}$. We know,
\begin{equation}\label{eq27}
 {R^{\left( {k + 1} \right)}} = {R^{\left( k \right)}} + {r^{\left( {k + 1} \right)}} \times \Pr \left\{ {{\rm{SU~transmittsin}}{{(k + 2)}^{th}}{\rm{channel}}} \right\}
\end{equation}

where ${r^{\left( k+1 \right)}}$ is defined in (\ref{eq23}) and $\Pr \left\{ {{\rm{SU~transmitts~in~}} {{(k+2)}^{th}}{\rm{channel}}} \right\}$ is equal to $\Pr \left\{ {{\rm{Number~of~HOs}} = k + 1} \right\}$, where based on independency of different channels, is equal to ${q_1} \times {q_2} \times  \cdots  \times {q_{k + 1}}$ , where $q_k$ is defined as (\ref{eq9}). Therefore,
\begin{equation}\label{eq28}
\begin{array}{r}
 {R^{\left( {k + 1} \right)}} = {R^{\left( k \right)}} + \left( {{C_1}{P_{k+2,1}}\left( {1 - {P_d}} \right) + {C_0}{P_{k+2,0}}\left( {1 - {P_{fa}}} \right)} \right) \times  \\
 {q_1}{q_2} \cdots {q_m}\left( {1 - \frac{{\tau  + \left( {k + 1} \right)\left( {\tau  + {\tau _{ho}}} \right)}}{T}} \right) \\
 \end{array}
\end{equation}

which leads to (\ref{eq12}) for $k = \alpha $, i.e., ${R^{\left( k \right)}} = {R^{\left( \alpha  \right)}} = R$.

\bibliographystyle{IEEEtran}
\bibliography{IEEEabrv,Bibliogeraphy}
\begin{table*}[h]
  \centering
    \caption{Simulation Parameters}\label{table1}
\begin{tabular}{|c|c|c|c|c|c|c|c|c|}
  \hline
  $P_d^{\min }$ & $P_{fa}^{\max }$ & $f_s$ (MHz) & $\gamma$  (dB) & Noise Spectral density & $T$  (ms) & $\tau_{ho}$  (ms)  & $N_p$ & $\frac{{{C_1}}}{{{C_0}}}$ \\
  \hline
  0.9 & 0.1 & 6 & -20 & -174 dBm/Hz & 100 & 0.1 & 10 & 0.1 \\
  \hline
 \end{tabular}
 \end{table*}

\end{document}